\documentclass[12pt]{iopart}
\usepackage{epsfig,graphicx,tabularx}
\usepackage{psfrag}
\usepackage{graphicx}
\usepackage{graphics}
\usepackage{epsfig}
\usepackage{bm}
\usepackage{verbatim,color,ulem}

\newcommand{\beq}{\begin{equation}}
\newcommand{\eeq}{\end{equation}}
\newcommand{\beqa}{\begin{eqnarray}}
\newcommand{\eeqa}{\end{eqnarray}}

\newcommand{\be}{\begin{equation}}
\newcommand{\ee}{\end{equation}}
\newcommand{\bea}{\begin{eqnarray}}
\newcommand{\eea}{\end{eqnarray}}

\begin{document}

\title{Itinerant ferromagnetism of two-dimensional repulsive 
fermions with Rabi coupling}

\author{V. Penna}
\address{Dipartimento di Fisica, Politecnico di Torino, 
and CNISM, Corso Duca degli Abruzzi 24, 
I-10129 Torino, Italy} 
\author{L. Salasnich}
\address{Dipartimento di Fisica e Astronomia ``Galileo Galilei'' 
and CNISM, Universit\`a di Padova, Via Marzolo 8, I-35131 Padova, Italy}
\address{Istituto Nazionale di Ottica (INO) del Consiglio Nazionale 
delle Ricerche (CNR), Via Nello Carrara 1, I-50019 Sesto Fiorentino, Italy}

\date{\today}

\begin{abstract}
We study a two-dimensional fermionic cloud of repulsive 
alkali-metal atoms characterized by two hyperfine states 
which are Rabi coupled. 
Within a variational Hartree-Fock scheme, 
we calculate analytically the ground-state energy of the system. 
Then we determine the conditions under which there is a quantum phase transition
with spontaneous symmetry breaking from a spin-balanced configuration to 
a spin-polarized one, an effect known as itinerant ferromagnetism. 
Interestingly, we find that the transition appears when the 
interaction energy per particle exceedes both the kinetic energy per particle 
and the Rabi coupling energy. The itinerant ferromagnetism of 
the polarized phase is analyzed, obtaining the population 
imbalance as a function of interaction strength, 
Rabi coupling, and number density. Finally, the inclusion 
of a external harmonic confinement is investigated by adopting 
the local density approximation. We predict that 
a single atomic cloud can display population imbalance near the center 
of the trap and a fully balanced configuration at the periphery. 
\end{abstract}


\maketitle

\section{Introduction}

Recently, artificial spin-orbit and Rabi couplings have been implemented 
by means of
counterpropagating laser beams in bosonic \cite{so-bose,so-bose2} and fermionic 
\cite{so-fermi1,so-fermi2} atomic gases.  
These laser beams couple two internal hyperfine states of the atom by 
a stimulated two-photon 
Raman transition \cite{so-bose,so-bose2,so-fermi1,so-fermi2}. Triggered by 
these remarkable experiments, 
in the last few years a large number of theoretical papers have analyzed the 
spin-orbit
effects with Rashba \cite{rashba} and Dresselhaus \cite{dress} terms in 
Bose-Einstein condensates
\cite{stringa1,stringa2,burrello,so-solitons,so-bright1,so-bright2,so-bright3} 
and also in the BCS-BEC crossover of superfluid fermions 
\cite{shenoy,gong,hu,iskin,dms,sa,chen,zhou2,yangwan}. Very recently, 
the Rashba spin-orbit coupling in a two-dimensional (2D) repulsive Fermi gas 
has been investigated in \cite{sala-previous,gigli}, where  
the density of states is quite simple and analytical results can be 
obtained. We stress that 2D 
quantum systems show peculiar physical properties and are crucial for 
technological applications: high-temperature superconductivity 
is attributed to materials characterized by a 2D-like transport \cite{htc}, 
and, more generally, superconductor and oxide interfaces 
containing 2D electron gas are of paramount importance 
for contemporary electronics \cite{interface}. 

The Rabi coupling of hyperfine states of atoms is now a 
common tool for experimental and theoretical investigations 
involving multi-component gases. Some examples are: 
the control of the population of the hyperfine levels 
\cite{lewensteinbook,steck}, the formation of localized structures 
\cite{horstman2010}, and the mixing-demixing dynamics of 
Bose-Einstein condensates \cite{ober}. 
It is particularly interesting to study how the Rabi coupling 
affects the equilibrium properties of an atomic two-dimensional repulsive
Fermi gas, and, more specifically, if the Rabi coupling can help a gas of
spin-up and spin-down fermions to become ferromagnetic, thus 
determining the itinerant ferromagnetism proposed in \cite{itinero1}.
The repulsive interaction induces the well-known Stoner 
instability \cite{stoner}
above a critical strength. Nevertheless, in the absence of the Rabi coupling
this instability is expected to produce phase separation 
rather than spin flip  \cite{itinero0}. Generally speaking, 
itinerant ferromagnetism is signaled by the spontaneous 
appearance of local spin imbalance, 
but this itself doesn't require spin flip mechanisms. 
A thorough investigation of this 
instability critical strength has been developed in 
\cite{itinero2,itinero3,itinero4,itinero5}. 
The observation of itinerant ferromagnetism in ultracold atoms in 3D 
is complicated by the presence of three-body losses \cite{sanner}, 
which are however expected to be less important in reduced 
dimensions \cite{petrov}. 
As noted in Ref. \cite{massi}, the itinerant ferromagnetism is a key effect to
get a deeper insight in the physics of systems such as metals, quark liquids in
neutron stars. Moreover, it is still debated whether homogeneous electron 
systems can reach a fully ferromagnetic state. We stress that very recently 
the observation of the ferromagnetic instability has been reported 
in a binary spin-mixture of ultracold $^6$Li atoms \cite{roati}. 

In this paper we study a Rabi-coupled fermionic gas 
of repulsive alkali-metal atoms trapped in a quasi two-dimensional 
configuration, where the effects of the third direction are fully frozen 
due to a strong external confinement in that direction 
\cite{sala-fermi-reduce}. Itinerant 
ferromagnetism in a trapped repulsive 2D Fermi system, but 
without Rabi coupling, has been investigated both 
analytically \cite{conduit0,china} 
and numerically \cite{conduit1}. The fermionic atoms are characterized by 
two hyperfine internal states which can be modelled as two spin components. 
Here we investigate the ground-state properties of the quantum gas by using 
the Hartree-Fock method in the form of a mean-field approximation for 
operator products, where the population imbalance is a variational parameter. 
In this way we calculate analytically the conditions under which there 
is a quantum phase transition 
from a spin-balanced to a spin-polarized  configuration. 
This phase transition features a spontaneous symmetry breaking of the 
fermion polarization (population imbalance) between
two degenerate values. The behavior of the population imbalance is determined 
as a function of the system parameters. We also consider the inclusion of an 
external harmonic potential investigating non-trivial effects 
caused by the space-dependent confinement on the polarization of the atomic 
cloud. 

\section{Model Hamiltonian}

The many-body Hamiltonian of the 2D fermionic atomic gas 
including contact interactions of strength $g$ and Rabi coupling 
of frequency $\Omega$ reads 
\beqa
{\hat H} 
=
\int d^2{\bf r} \Bigl [\; \sum_{\sigma=\uparrow,\downarrow} 
{\hat \psi}_{\sigma}^+ \left( -\frac{\hbar^2}{2m}\nabla^2 \right) \, 
{\hat \psi}_{\sigma}  
&+&  
g \ 
{\hat \psi}_{\uparrow}^+ {\hat \psi}_{\downarrow}^+ {\hat \psi}_{\downarrow} 
{\hat \psi}_{\uparrow}
\nonumber 
\\ 
&+& \frac{\hbar \Omega}{2} \left(  {\hat \psi}_{\uparrow}^+ 
{\hat \psi}_{\downarrow} + 
{\hat \psi}_{\downarrow}^+ {\hat \psi}_{\uparrow}  \right)  \Bigr ], 
\label{H2}
\eeqa
where ${\hat \psi}_{\sigma}({\bf r})$ is the field operator which destroys 
a fermion of spin $\sigma$ at position ${\bf r}$. 
It is important to stress that, due to the presence of the Rabi coupling, 
the total number  
\beq 
{\hat N} = \int d^2{\bf r} \left( 
{\hat \psi}_{\uparrow}^+({\bf r}) {\hat \psi}_{\uparrow}({\bf r}) + 
{\hat \psi}_{\downarrow}^+({\bf r}) {\hat \psi}_{\downarrow}({\bf r})
\right) ,
\eeq
is a constant of motion, while the relative numbers ${\hat N}_{\uparrow}$ 
and ${\hat N}_{\downarrow}$ are not. 

Applying the mean-field approximation for operator products 
(see, e.g., \cite{MF1}, \cite{MF2}) to
\beq 
{\hat \psi}_{\uparrow}^+ {\hat \psi}_{\downarrow}^+ {\hat \psi}_{\downarrow} 
{\hat \psi}_{\uparrow} \simeq 
n_{\downarrow} \ {\hat \psi}_{\uparrow}^+ {\hat \psi}_{\uparrow} 
+ n_{\uparrow} \ {\hat \psi}_{\downarrow}^+ {\hat \psi}_{\downarrow} 
- n_{\uparrow} \, n_{\downarrow} ,
\eeq
with 
$n_{\sigma} = \langle {\hat n}_{\sigma}\rangle=\langle 
{\hat \psi}_{\sigma}^+ {\hat \psi}_{\sigma} \rangle$ 
($\sigma=\uparrow,\downarrow$), enables us to write the mean-field 
many-body Hamiltonian as 
\beq 
{\hat H} =  \int d^2{\bf r} \left [ \left( {\hat \psi}_{\uparrow}^+ ,  
{\hat \psi}_{\downarrow}^+\right) 
\ {\cal H} \ 
\left( \begin{array}{c} 
{\hat \psi}_{\uparrow} \\
{\hat \psi}_{\downarrow}
\end{array} \right) \right ] - {gn^2\over 4} (1-\zeta^2) L^2 ,
\label{HMF} 
\eeq
where $L^2$ is the area of the 2D system and 
${\cal H}$ is the single-particle matrix Hamiltonian
\beq 
{\cal H} =  
\left( \begin{array}{c c } 
-\frac{\hbar^2}{2m}\nabla^2  + \frac{gn}{2} (1+\zeta ) 
& \frac{\hbar \Omega}{2} \\ 
\frac{\hbar \Omega}{2} & -\frac{\hbar^2}{2m}\nabla^2 + 
\frac{gn}{2} (1-\zeta ) 
\\ 
\end{array} \right)  
\label{HSP}
\eeq
with the average total number density and the population imbalance
given by
\beq 
n=n_{\uparrow} +n_{\downarrow} \, ,\qquad
\zeta ={n_{\downarrow }-n_{\uparrow}\over n}  ,
\eeq
respectively. 
Clearly, at fixed total density ${n}$, one finds that $\zeta\in [-1,1]$.  
It is important to stress that, within our Hartree-Fock scheme, 
$\zeta$ is a variational parameter which must be determined by minimizing 
the energy of the system. 
By using the Pauli matrices  $\sigma_z$ and $\sigma_x$ such that 
$[ \sigma_a, \sigma_b ] = i\epsilon_{abc} \sigma_c$, with indexes 
$a,b, c= x,y,z$, the single-particle Hamiltonian (\ref{HSP}) takes the form
$$
{\cal H} = \left(  -\frac{\hbar^2}{2m}\nabla^2  
+ \frac{gn}{2} \right) \ {\cal I}  + \frac{gn}{2} \ \zeta \ \sigma_z 
+ \frac{\hbar \Omega}{2} \sigma_x \; .
$$
The latter can be diagonalized exactly \cite{diag}, and one finds 
\beq 
{\cal H} |{\bf k},s\rangle = E_{{\bf k},s} |{\bf k},s \rangle\; ,\quad
\eeq
where the eigenvalue 
\beq 
E_{{\bf k},s} =  
\frac{ \hbar^2k^2}{2m} + \alpha_s 
\label{energia}
\eeq
depends on the two-dimensional wavevector  ${\bf k}$, the index $s=-1,+1$ 
is the eigenvalue of $\sigma_z$, and 
\beq 
\alpha_s =  \frac{g}{2} {n} + {s\over 2} \; \sqrt{ {g^2n^2} \zeta^2  
+ {\hbar^2 \Omega^2} } \, 
\label{energia-pezzetto}
\eeq
is the contribution to the single-particle energy due to the repulsive 
interaction of strength $g$ ($g>0$) and the Rabi coupling of frequency 
$\Omega$. 
The corresponding eigenstates are given by
$$
|{\bf k},s \rangle=  \frac{e^{i {\bf k }\cdot {\bf r}}}{L}  \; S |s \rangle ,
$$
where $\sigma_z |s\rangle  = s \; |s\rangle $,  
${\vec p}\; |{\bf k}, s \rangle = \hbar {\bf k } \;|{\bf k},s \rangle$ and 
$S = \exp ({i \phi \sigma_y/2} )$,
with ${\rm tg} \phi = {\hbar \Omega}/{gn}$, is the transformation 
taking $\cal H$ into the diagonal form. 
It follows that the mean-field many-body Hamiltonian can be written as 
\beq 
{\hat H} = - {gn^2\over 4} (1-\zeta^2) L^2 + 
\sum_{\bf k} \sum_{s=-1,1} E_{{\bf k},s} 
\ {\hat b}_{{\bf k},s}^{+} \ {\hat b}_{{\bf k},s} , 
\label{h-mf}
\eeq 
where ${\hat b}_{{\bf k},s}$ and ${\hat b}_{{\bf k},s}^{+}$ are ladder 
operators which destroy and create a fermion in the single-particle 
state $|{\bf k},s\rangle$. 

\section{Ground-state properties} 

By implementing the continuum limit
$\sum_{\bf k} \to L^2\int d^2{\bf k}/(2\pi)^2$, the average total number 
density $n=N/L^2$ of the fermionic system is found to be 
\beq 
n = \sum_{s=-1,1} 
\int {d^2{\bf k}\over (2\pi)^2} \langle {\hat b}_{{\bf k},s}^{+} 
\ {\hat b}_{{\bf k},s} \rangle , 
\label{number}
\eeq
while the average internal-energy density ${\cal E}=E/L^2$ reads 
\beq 
{\cal E} = - {gn^2\over 4} (1-\zeta^2) + 
\sum_{s=-1,1} 
\int {d^2{\bf k}\over (2\pi)^2} E_{{\bf k},s}
\langle {\hat b}_{{\bf k},s}^{+} 
\ {\hat b}_{{\bf k},s} \rangle  . 
\label{energy}
\eeq
Moreover, at zero temperature one can write 
\beq 
\langle {\hat b}_{{\bf k},s}^{+} 
\ {\hat b}_{{\bf k},s} \rangle = 
\Theta\left(\mu -E_{{\bf k},s}  \right) ,  
\label{sembrafacile1}
\eeq
where $\Theta(x)$ is the Heaviside step function and $\mu$ is the 
zero-temperature chemical potential, namely, the Fermi energy of the 
interacting system. Notice that $\mu$ is fixed by the conservation 
of the total number of fermions. 
Then, by using Eq. (\ref{sembrafacile1}), from Eqs. (\ref{number}) 
and (\ref{energy}) we find 
\beq   
n = {1 \over 4\pi} \left({2m\over \hbar^2}\right) 
\Big[ (\mu - \alpha_{-1}) \ \Theta(\mu - \alpha_{-1}) + (\mu - \alpha_{+1}) 
\ \Theta(\mu - \alpha_{+1})  \Big]
\label{number1}
\eeq
and 
\beqa 
{\cal E} = &-& {gn^2\over 4} (1-\zeta^2)  + {1 \over 8\pi} 
\left( {2m\over \hbar^2}\right)  \Big[ (\mu^2 - \alpha^2_{-1}) \ 
\Theta(\mu - \alpha_{-1})
\nonumber
\\
&+& (\mu^2 - \alpha^2_{+1})  \ \Theta(\mu - \alpha_{+1}) \Big]  . 
\label{energy1}
\eeqa
Clearly, if $\mu \leq \alpha_{-}$ there are no solutions. 
Let us now consider the remaining cases $\mu < \alpha_{+}$ and 
$\alpha_{+}\leq \mu$. 

\begin{center}
{\bf Regime $\mu < \alpha_{+}$}
\end{center}

From Eqs. (\ref{energia-pezzetto}),  (\ref{number1}) and (\ref{energy1}), 
under the condition $\mu < \alpha_{+1}$, we obtain 
\beq
\mu = {4\pi \hbar^2\over 2m}\; n + {1\over 2} gn 
- {1\over 2} \sqrt{g^2n^2\zeta^2+\hbar^2\Omega^2} 
\label{muio}
\eeq
and also  
\beq
{\cal E} = - {gn^2\over 4} (1-\zeta^2) 
+ \frac{n}{2} \left [ \left (  g  + \frac{4\pi \hbar^2}{2m} \right )\; n 
-  \sqrt{{g^2n^2} \zeta^2 + {\hbar^2\Omega^2}} \; \right ],
\label{buio}
\eeq
where Eq. (\ref{muio}) has been used to express ${\cal E}$ in terms of $n$ 
instead of $\mu$. This average energy density ${\cal E}$ 
is a function of the population imbalance $\zeta$, which is our 
variational parameter. For the sake of simplicity we introduce 
the characteristic energies 
$$
E_{\Omega} = \hbar \Omega \; , \qquad
E_{int} = g n \; , 
\qquad E_{kin} = {4\pi \hbar^2 n\over 2m} \; . 
$$
The minimum of ${\cal E}$ with respect to $\zeta$ 
is easily found from the condition 
${\partial {\cal E}/\partial \zeta} = 0$
which, written in terms of $E_{int}$ and $E_{\Omega}$, gives 
\beq 
\frac{n}{2} E_{int} \zeta \ \left ( 1 - 
\frac{ E_{int} }{ \sqrt{ E^2_{int} \zeta^2+ E^2_{\Omega} } } \right ) = 0 .
\eeq
Consequently, one has two cases: $\zeta = 0$ for $E_{int} \le E_{\Omega}$, and 
$\zeta = \pm \sqrt{ 1 -{E_{\Omega}^2}/{E^2_{int} }  }$ for 
$E_{\Omega} <E_{int}$. 
In the second case, the solution $\zeta =0$ describes a maximum separating 
the two minima. 
This scenario is completed by taking into account the 
condition $\mu < \alpha_{+}$ characterizing the present regime, 
with $\mu$ given by Eq. (\ref{muio}), finding 
\beq 
E_{kin} \; < \sqrt{ E^2_{int} \zeta^2  +E^2_{\Omega}} .
\eeq
Then, the two cases described above can be detailed as follows.  

\noindent 
{\bf Condition A}: For $E_{int}  \le E_{\Omega}$ and 
$E_{kin} < E_{\Omega}$ the population imbalance is 
\beq 
\zeta_A = 0 , 
\eeq
and the corresponding chemical potential and energy density are given by
%
\beq 
\mu_A = \frac{1}{2} E_{int}  + E_{kin} - \frac{1}{2} E_{\Omega} ,  
\qquad
{\cal E}_A = {n \over 4} E_{int}  
+ {n \over 2} \left ( E_{kin} - E_{\Omega} \right ) . 
\eeq

\noindent
{\bf Condition B}: For $E_{\Omega} < E_{int}$ and $E_{kin} < E_{int}$ the population imbalance is 
\beq 
\zeta_B = \pm \sqrt{ 1 -\left(\frac{E_{\Omega}}{E_{int}}\right)^2  } ,  
\label{mainresults}
\eeq
which shows the double degeneracy of the ground state and entails
a spontaneous symmetry breaking, while
%
\beq 
\mu_B = E_{kin} , \qquad 
{\cal E}_B =  {n \over 2} \left ( E_{kin} - \frac{E^2_{\Omega}}{2E_{int}} \right )
\eeq
represent the chemical potential and energy density, respectively.
The results under the condition B) show explicitly that there is 
population imbalance if the interaction energy  
per particle $E_{int}$ is larger than both the kinetic term 
$E_{kin}$ (proportional to the kinetic energy per particle 
$\pi \hbar^2 n/(2m)$) and the Rabi energy $E_{\Omega}$. 
This is a clear example of Stoner instability \cite{stoner}, where a 
sufficiently large repulsion between fermions makes the uniform 
and balanced system unstable. In this case, due to the presence 
of Rabi coupling, the system becomes polarized being either 
$n_{\uparrow} > n_{\downarrow}$ ($\zeta_B <0$) or $n_{\uparrow} < n_{\downarrow}$ 
($\zeta_B >0$). 

\begin{center}
{\bf Regime $\alpha_{+}\leq \mu$} 
\end{center}

From Eqs. (\ref{energia-pezzetto}), (\ref{number1}) and (\ref{energy1}), 
under the condition $\alpha_{+1}\leq \mu$, we obtain 
\beq
\mu = {1 \over 2} \left ( E_{kin} + E_{int} \right )
\label{muio2}
\eeq
and the ground-state energy
%
\beq 
{\cal E} = {n \over 4} \left [ E_{int} \zeta^2 \left( 1 - \frac{E_{int}}{E_{\Omega}} \right) 
+  2 E_{int} + E_{kin}  - \frac{ E^2_{\Omega}}{E_{kin}} \right ] \;   
\label{buio2}
\eeq
by using Eq. (\ref{muio2}) to express ${\cal E}$ in terms of $n$ 
instead of $\mu$. Also in this case the average energy density ${\cal E}$ 
is a function of the population imbalance $\zeta$, which is our 
variational parameter. However, the functional dependence of (\ref{buio2}) 
on $\zeta$ is quite different with respect to (\ref{buio}).  

Finding the minimum of ${\cal E}$, given by Eq. (\ref{buio2}), 
with respect to $\zeta$ gives two cases: $\zeta = 0$ for 
$ E_{int} < E_{kin}$, and $\zeta = \pm 1$ for 
$ E_{kin} <E_{int}$. Again, one must include 
the condition $\alpha_{+} \leq \mu$, with $\mu$ given by Eq. (\ref{muio2}), 
obtaining 
%
\beq 
\sqrt{ \zeta^2 E^2_{int}  + E^2_{\Omega} } \leq  E_{kin} .
\label{mahmah}
\eeq
One easily discovers that the second case described above ($\zeta=\pm 1$ 
for $ E_{kin} <E_{int} $) is incompatible with (\ref{mahmah}) and it must be excluded.  By taking 
into account (\ref{mahmah}), the remaining case characterized by $E_{int}< E_{kin}$) can be 
detailed as follows.  

\noindent
{\bf Condition C}: For $E_{int} \leq E_{kin}$ and $E_{\Omega} \leq E_{kin} $ the population 
imbalance  
\beq 
\zeta_C = 0 , 
\eeq
entails 
\beq 
\mu_C = \frac{1}{2} \Bigl ( E_{int}  + E_{kin} \Bigr ) , \qquad  
{\cal E}_C = {n \over 4} \left ( E_{int}   + E_{kin}  - \frac{E^2_{\Omega}}{E_{kin}} \right )
\eeq
representing the chemical potential and energy density, respectively, 
of this case.

\begin{figure}[tbp]
\centering
\epsfig{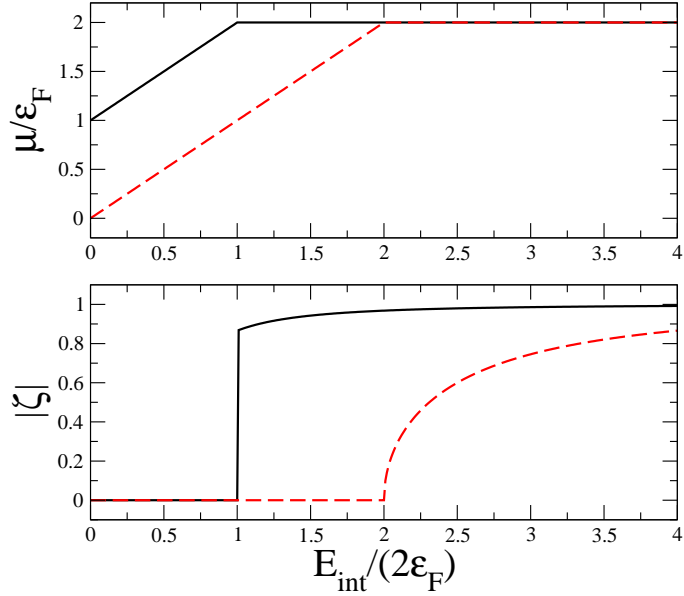}
\caption{Upper panel: adimensional chemical potential $\mu/\epsilon_F$ as 
a function of the adimensional interaction strength $E_{int}/(2\epsilon_F)$  
for two values of adimensional Rabi energy  $E_{\Omega}/(2\epsilon_F)$: 
$1/2$ (solid line) and $2$ (dashed line). 
Lower panel: absolute value $|\zeta|$ of the population imbalance  as a 
function of the adimensional interaction strength $E_{int}/(2\epsilon_F)$ 
for the same two values of adimensional 
Rabi energy $E_{\Omega}/(2\epsilon_F)$.}
\label{fig1} 
\end{figure}

The analysis so far developed clearly shows that only under the condition B) 
there is itinerant ferromagnetism in the two-dimensional repulsive 
Fermi gas. The condition B) means that the interaction energy 
per particle $E_{int}$ must be larger than both the kinetic energy 
per particle $E_{kin}$ and the Rabi energy $E_{\Omega}$. 
To summarize, this result is convenient to introduce 
the Fermi energy $\epsilon_F$ of our 2D fermionic system in the absence of 
interaction and Rabi coupling, that is given by 
\beq 
\epsilon_F = { \hbar^2\over m} \; \pi n \; .  
\eeq
Taking into account the conditions A, B, C described in the previous Section,  
the chemical potential of the system in the presence of interaction 
and Rabi coupling can be then written as 
\beq 
\mu = 
\left\{ \begin{array}{ll}
\epsilon_F + {1\over 2} E_{int} & \mbox{  for  } E_{int} \leq 2 \epsilon_F \\
2\epsilon_F & \mbox{  for  } E_{int} > 2 \epsilon_F
\end{array} \right. 
\label{mico1}
\eeq
under the condition $E_{\Omega} \leq 2\epsilon_F$, and 
\beq
\mu = \left\{ \begin{array}{ll}
2\epsilon_F - {1\over 2} \left( E_{\Omega}  - E_{int} \right)  
& \mbox{  for  } E_{int} \leq E_{\Omega} 
\\ 
2\epsilon_F & \mbox{  for  } E_{int} > E_{\Omega} 
\end{array} \right. 
\label{mico2}
\eeq
under the condition $E_{\Omega} > 2\epsilon_F$. 
In the upper panel of Fig. \ref{fig1} we report the 
adimensional chemical potential $\mu/\epsilon_F$ 
as a function of the adimensional interaction strength $E_{int} /(2\epsilon_F)$ 
for two values of adimensional Rabi energy $E_{\Omega} /(2\epsilon_F)$. 
The figure clearly shows that at the critical strength there is the derivative 
of the chemical potential changes slope. 

The region where $\mu=2\epsilon_F$ corresponds 
to the condition B: the system becomes spin-polarized. 
In the lower panel of Fig. \ref{fig1} we plot 
the population imbalance $|\zeta|$ as a function of the adimensional 
interaction strength $gn/(2\epsilon_F)$ for two values of 
adimensional Rabi energy $E_{\Omega}/(2\epsilon_F)$. 
As shown in the figure, the population imbalance $\zeta$, given by 
Eq. (\ref{mainresults}), decreases by increasing the Rabi frequency $\Omega$. 
This result is consistent with previous two-dimensional 
calculations \cite{itinero2,itinero3,itinero4,itinero5}
which suggest, in the absence of Rabi coupling, a jump from 
$\zeta=0$ to $\zeta=\pm 1$ at the critical 
strength $g_c= 4\pi \hbar^2 /(2m)= E_{kin}/n$. 
Notice that this jump can be softened also by beyond-mean-field quantum 
effects \cite{itinero2} or spin-orbit couplings \cite{itinero4}. 
Our results on the order parameter $\zeta$, and specifically 
the lower panel of Fig. \ref{fig1}, signal a first-order phase 
transition if $E_{\Omega} / (2\epsilon_F)<1$ and a second-order 
phase transition if $E_{\Omega}/( 2\epsilon_F)>1$.

\section{Discussion and inclusion of harmonic confinement} 

Up to now we have considered a 2D homogeneous fermionic system. 
Here we discuss the effect of an external hamonic confinement 
\beq 
U({\bf r}) = {1\over 2} m \omega^2 (x^2 +y^2) 
\eeq
on the properties of the 2D system. We adopt the local density 
approximation \cite{lip}:  
\beq 
{\bar \mu} = \mu [n({\bf r})] + U({\bf r}) \; , 
\label{lda}
\eeq
where ${\bar \mu}$ is the chemical potential of the non uniform 2D system, 
$n({\bf r})=n_{\uparrow}({\bf r})+n_{\downarrow}({\bf r})$ is the local 
number density with 
\beq 
N = \int d^2{\bf r} \ n({\bf r}) 
\eeq 
the total number of fermions, and  $\mu[n]$ is the local chemical 
potential given by Eqs. (\ref{mico1}) and (\ref{mico2}). 
\begin{figure}[tbp]
\centering
\epsfig{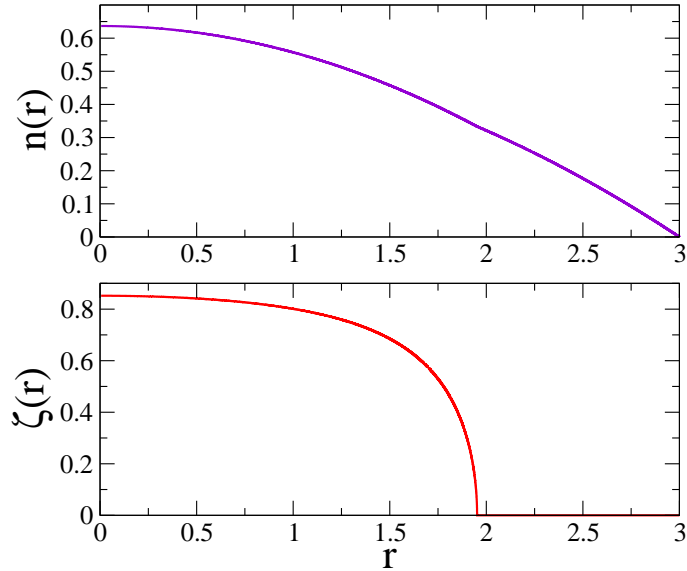}
\caption{Total density profile $n(r)$ 
and population imbalance profile $\zeta(r)$ of the 2D Fermi gas under 
harmonic confinement of frequency $\omega$. 
We set $\hbar=m=\omega=1$ and choose $\Omega=1$, 
$g=3$, and $\bar{\mu}=4$. The lower panel shows spin-flip (population 
imbalance) near the center of the trap while the periphery remains balanced.} 
\label{fig2} 
\end{figure}
By using $r=\sqrt{x^2+y^2}$ Eq. (\ref{lda}) can be written as 
\beq 
r = {1\over \omega} \sqrt{2 \left( \bar{\mu} - \mu[n] \right)} 
\eeq
which gives the radial coordinate $r$ as a function of the number density $n$. 
This formula can be easily implemented numerically to determine 
the density profile $n(r)$, the local population imbalance 
%
\beq 
\zeta(r) = \left\{ \begin{array}{ll} 
0 & \mbox{  for  } n(r) \leq {E_{\Omega} \over g} \\ 
\sqrt{1 - {E^2_{\Omega} / (g^2 n^2(r)) }} 
& \mbox{  for  } n(r) > {E_{\Omega} \over g} 
\end{array} \right. \; , 
\eeq
and the local densities of fermions with spin up and spin down:  
\beq 
n_{\downarrow}(r) = {1\over 2} n(r) \Bigl ( 1 + \zeta(r) \Bigr ) \; , 
\quad\quad\quad  
n_{\uparrow}(r)= {1\over 2} n(r) \Bigl ( 1 - \zeta(r) \Bigr ) \; . 
\eeq

As an example, in Fig. \ref{fig2} we report the total number 
density profile $n(r)$ (upper panel) 
and the population imbalance profile $\zeta(r)$ (lower panel) 
with a simple choice of the parameters which ensures that 
$n(0)> E_{\Omega}/g$. This condition is crucial to produce an atomic 
cloud with population imbalance. Note that the appearance of a 
non-zero population imbalance implies a spontaneous 
symmetry breaking of the ground-state with respect to the 
choice $\zeta$ or $-\zeta$.  

\begin{figure}[tbp]
\centering
\epsfig{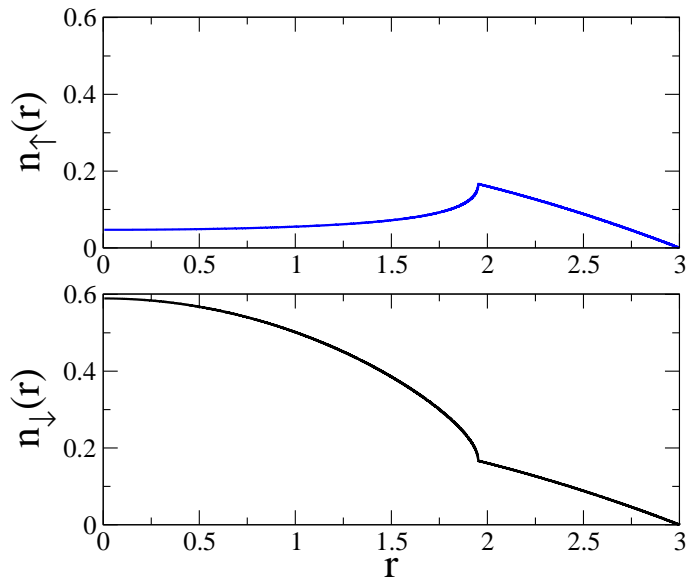}
\caption{Local densities $n_{\downarrow}(r)$ 
and $n_{\uparrow}(r)$ of the 2D Fermi gas under harmonic confinement 
of frequency $\omega$. We set $\hbar=m=\omega=1$ and choose $\Omega=1$, 
$g=3$, and $\bar{\mu}=4$. The panels clearly show the enhancement 
of the number of atoms with spin down near the center of the trap, 
while near the surface the system remains balanced.} 
\label{fig3} 
\end{figure}

In Fig. \ref{fig3} we plot the corresponding local 
densities $n_{\downarrow}(r)$ and $n_{\uparrow}(r)$. 
The figures clearly show that the atomic cloud is characterized 
by population imbalance near the center of the trap ($r=0$) where 
the total number density is larger than $E_{\Omega}/g$. 
Instead, at the periphery (near the surface) of the atomic cloud 
the gas is fully balanced. We emphasize that setting 
$\hbar=m=\omega=1$ (as done in Fig. \ref{fig2} and Fig. \ref{fig3}) 
amounts to using harmonic-trap units: 
energies in units of $\hbar\omega$ and lengths in units 
of $a_{H}=\sqrt{\hbar/(m\omega)}$, that is 
the characteristic length of harmonic confinement. 

In the experiments with ultracold atomic clouds having a 
quasi-2D disk-shaped configuration on the $(x,y)$ plane, 
one finds typically $a_{H} \simeq 100$ $\mu$m, while 
the 2D interaction strength $g$ reads $g= 4 (\hbar \omega)a_s a_H^2/a_z$ 
with $a_s$ the 3D s-wave scattering length and $a_z$ the characteristic 
length of the confinement along the $z$ axis. 
Remarkably, in current experiments the 3D s-wave scattering length $a_s$ 
can be modified by using an external magnetic field 
(Fano-Feshbach resonance technique) and consequently one can easily 
move the system from a weakly-interacting to a strongly-interacting regime. 

\section{Conclusions} 

In this paper we have shown how the non-trivial interplay 
among Pauli exclusion principle, repulsive interaction, 
and Rabi coupling can induce 
itinerant ferromagnetism in two-dimensional repulsive Fermi gases. 
In particular, we have analytically found that for a homogeneous 
2D fermionic system there is polarization (i.e., itinerant ferromagnetism) 
when the interaction energy per particle is larger than both the kinetic energy 
per particle and the Rabi energy. It is important to stress that 
the itinerant ferromagnetism is certainly driven by the Stoner 
instability \cite{stoner}: a sufficiently large repulsion 
between fermions make the uniform and balanced system unstable. 
However, as we have shown in this paper, it is the presence of Rabi coupling 
that allows the phenomenon of spin flip. In fact, in the absence 
of Rabi coupling or other spin-dependent mechanisms, the Stoner 
instability implies phase separation and not spin flip. 
Similar effects are expected in bosonic mixtures \cite{ober,enna}. 
Here we have adopted a Hartree-Fock mean-field approach. 
On the basis of previous results obtained in the absence of Rabi coupling 
in 2D and 3D \cite{conduit0,conduit1,pilati}, we expect that beyond-meand-field 
quantum fluctuations can slightly reduce the critical 
strength of Stoner instability. 

In the last part of the paper we have considered 
the inclusion of an external harmonic 
potential, which is the simplest trapping configuration for experiments 
with ultracold alkali-metal atoms. In this case, we have predicted 
a remarkable effect we expect to be accessible in the near-future experiments: 
for a sufficiently large number of fermions, such that the 
number density at the center of the trap exceeds a critical value, 
the 2D fermionic gas is characterized by population imbalance near the center 
of the trap and by a fully balanced configuration near the surface.

\section*{Acknowledgements}

The authors thank Prof. Flavio Toigo for enlightening discussions and 
suggestions. LS acknowledges for partial support the 2016 BIRD project 
"Superfluid properties of Fermi gases in optical potentials" of the 
University of Padova.

\end{document}